\documentclass[aps,pra,twocolumn,showpacs,superscriptaddress,groupedaddress]{revtex4}
\usepackage{graphicx}% Include figure files
\usepackage{dcolumn}% Align table columns on decimal point
\usepackage{bm}% bold math

 \usepackage{CJK}
\usepackage{amsmath} 
\usepackage{esint}
\usepackage{amsfonts}
\usepackage{mathrsfs} 
\usepackage{tabularx}  % ???
\usepackage{booktabs}  % ???????
\usepackage{multirow}

\usepackage{braket}
\usepackage{color}
\usepackage[colorlinks,
linkcolor=blue,
anchorcolor=blue,
citecolor=blue]{hyperref}

\begin{document}
\title{Module for arbitrary controlled rotation in gate-based quantum algorithms}

\author{Shilu Yan}
\address{School of Automation Science and Engineering, South China University of Technology, Guangzhou 510640, China}
\author{Tong Dou}
\address{School of Automation Science and Engineering, South China University of Technology, Guangzhou 510640, China}
\author{Runqiu Shu}
\address{School of Automation Science and Engineering, South China University of Technology, Guangzhou 510640, China}
\author{Wei Cui}
\email{aucuiwei@scut.edu.cn}
\address{School of Automation Science and Engineering, South China University of Technology, Guangzhou 510640, China}
\affiliation{}

\date{\today}

\begin{abstract}
To assess whether a gate-based quantum algorithm can be executed successfully on a noisy intermediate-scale quantum (NISQ) device, both complexity and actual value of quantum resources should be considered carefully. Based on quantum phase estimation, we implemente arbitrary controlled rotation of quantum algorithms with a proposed modular method. The proposed method is not limited to be used as a submodule of the HHL algorithm and can be applied to more general quantum machine learning algorithms. Compared with the polynomial-fitting function method, our method only requires the least ancillas and the least quantum gates to maintain the high fidelity of quantum algorithms. The method theoretically will not influence the acceleration of original algorithms. Numerical simulations illustrate the effectiveness of the proposed method. Furthermore, if the corresponding diagonal unitary matrix can be effectively decomposed, the method is also polynomial in time cost.

\end{abstract}

%\pacs{} 

\maketitle

%{\color{red}  text}

\section{\label{Sec1}Introduction}
Since Feynman showed his belief of that physics can be simulated with quantum computers\cite{feynman1982simulating}, quantum computing has developed over the last forty years. A variety of quantum algorithms that potentially give quantum speedups over classical computers have been proposed. Although building a universal quantum computer remains extremely difficult, researches underway at multiple technology companies, among IBM, Google, Rigetti, IonQ, Microsoft, Honeywell, D-wave and Origin Quantum, have led to a series of technological breakthroughs in building quantum computer systems. Nowadays, some platform has been provided for the public with access to cloud-based quantum computers.

Usually a complete gate-based quantum algorithm is modular, where each module has independent functions. Most gate-based quantum algorithms involve a module
\begin{eqnarray}
\ket{\lambda} \mapsto Cf(\lambda)\ket{\lambda},
\label{cr}
\end{eqnarray}
where $\ket{\lambda} \in (\mathbb{C}^{2})^{\otimes m}, $ C is a normalizing constant and $ f(\lambda) $ is an arbitrary function of $ \lambda $. The module can also be defined as quantum digital-to-analog conversion (QDAC) \cite{PhysRevA.99.012301}, or the oracle discussed in Sec.~\ref{Sec5}. Eq.~(\ref{cr}) can be accomplished by performing a theoretical controlled rotation
\begin{eqnarray} 
U_{cr}: \ket{0}\ket{\lambda} \mapsto Cf(\lambda)\ket{1}\ket{\lambda} + \sqrt{1-(Cf(\lambda))^{2}} \ket{0}\ket{\lambda},
\label{U_cr} 
\end{eqnarray}
measuring the first qubit, and repeating the procedure until the measurement results in $\ket{1}$. The famous example is the Harrow-Hassidim-Lloyd (HHL) algorithm \cite{PhysRevLett.103.150502}, which is widely applied in quantum machine learning to solve the classical standard linear system of equations. The HHL algorithm requires a typical controlled rotation with $ f(\lambda) = \lambda^{-1}$. To implement the typical controlled rotation, Ref.~\cite{Cao_2013} first used the Newton's iteration to approximate $ \lambda^{-1} $ and then evaluated the arcsine function by a bisection search. Ref.~\cite{cong2016quantum} implemented the inversion and inverse trigonometric function with the Taylor series and Maclaurin series, respectively. Refs.~\cite{wang2020quantum, Li_2020} calculated the rotation angular coefficients with evaluating inverse trigonometric function in a recursive way by a binary expansion method. More efficiently, Refs.~\cite{haner2018optimizing, vazquez2020enhancing} showed how to use piecewise polynomial approximation for arbitrary functions and provided a rigorous performance theory. 

However, these polynomial-fitting methods still use a large number of qubits to implement the controlled rotation. If more bits of precision are required, the number can easily stretch to several hundred qubits. This will severely limit the implementation efficiency of the entire quantum algorithm and even inhibit the acceleration. The implementation of HHL algorithms are facing the challenge of limited quantum resources. The core idea of reducing the cost of qubits is combining the function outputs and quantum circuit design, where the Refs.~\cite{PhysRevLett.93.130502, saeedi2013synthesis,shafaei2013reversible} can be seen as the different concrete implementations of this kind of method. Following the core idea, we propose a modular method to achieve arbitrary controlled rotation based on phase estimation. Under the preconditions of appling the original algorithm, our method theoretically will not influence the acceleration of the original algorithm. Compared with the polynomial-fitting function method, our method only requires the least ancillas and the least quantum gates in some way. Furthermore, if the corresponding diagonal unitary matrix can be effectively decomposed, our method is also polynomial in time cost. 

The paper is organized as follows. Section~\ref{Sec2} describes the proposed controlled rotation module based on phase estimation. In Section~\ref{Sec3}, we present different applications of the controlled rotation module. Section~\ref{Sec4} verifies our methods with numerical simulations. Section~\ref{Sec5} are analysis and discussions. Conclusions are presented in Section~\ref{Sec6}.

{\bfseries Notation}. The bold italic $ \boldsymbol {i} $ represents the imaginary unit of the complex field. $ FT^{\dagger} $ is the inverse quantum Fourier transform. {The $ R_{z}(\alpha) $ gate is defined as $ R_{z}(\alpha)= \left[
	\begin{matrix}
	1 & 0 \\
	0 & e^{2 \pi \boldsymbol {i} \alpha } 
	\end{matrix} \right] $.}

\section {\label{Sec2}The controlled roation module based on phase estimation}

\begin{figure*}[htbp]
	\centering
	\includegraphics[scale=0.6]{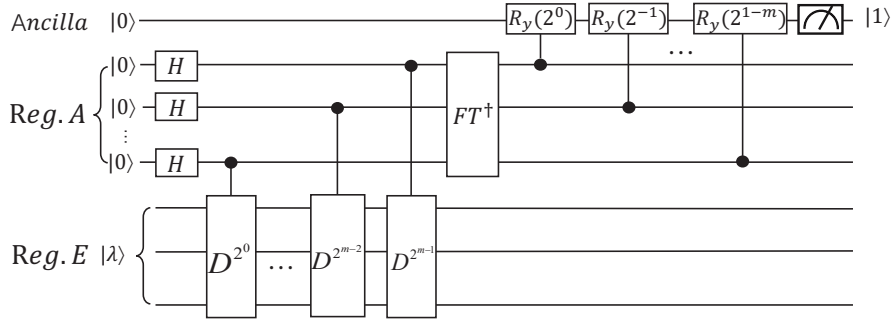}
	\caption{The controlled rotation module based on phase estimation.}
	\label{fig:method}
\end{figure*}

In the Eq.~(\ref{U_cr}), $\ket{\lambda}$ stores a binary representation of numbers, specifically, the eigenvalues of a matrix.  To find a circuit performing the transformation $ U_{cr} $, it can be divided into two parts: a unitary operator  
\begin{eqnarray}
U_{f}: \ket{\lambda}\ket{0}^{\otimes m} \mapsto \ket{\lambda}\ket{\dfrac{1}{2}arcsin(Cf(\lambda))}, 
{\label{U_{f}}}
\end{eqnarray}
and a standard controlled $ R_{y} $ rotation 
\begin{eqnarray}
R_{y}(2\theta) = \left[
\begin{matrix}
\sqrt {1 - {{\left( {Cf\left( \lambda  \right)} \right)}^2}} & - Cf\left( \lambda  \right) \\
Cf\left( \lambda  \right) & \sqrt {1 - {{\left( {Cf\left( \lambda  \right)} \right)}^2}} 
\end{matrix} \right],
{\label{R_{y}}}
\end{eqnarray}
where $ \theta = \dfrac{1}{2}{\arcsin \left( {Cf\left( \lambda  \right)} \right)} $. The quantum circuit for the proposed controlled rotation module is presented in Fig.~\ref{fig:method}. The unitary $ D $ is a diagonal operator, whose diagonal elements are $ e^{{2 \pi \boldsymbol {i}} \theta} $. Note that the unitary operator $ D $ has an eigenvector $ \ket{\lambda} $ with eigenvalue $ e^{{2 \pi \boldsymbol {i}} \theta} $, where the $ \theta $ is the most important factor in the generalization of the circuit. Eq.~(\ref{U_{f}}) can be seen as the procedure of phase estimation, which is used to approximate the eigenvalues of a discretized matrix and entangle the states encoding the eigenvalues with the corresponding eigenstates \cite{PhysRevA.54.4564}. 

The controlled rotation procedure uses two registers and an ancilla. Both registers contain $ m $ qubits with $ Reg.E $ in the state $ \ket{\lambda} $ obtained by previous step and $ Reg.A $ initially in $ \ket{0}^{\otimes m} $.  The number $ m  $ depends on two factors: one is the number of digits of accuracy in the estimation for $ \theta $, and the other is the probability of successful estimation \cite{nielsen2010quantum}. The circuit in Fig.~\ref{fig:method} begins by applying a Hadamard transform to the $ Reg.A $, followed by an application of controlled-$ D^{2^{i}} $ operations, where $ i = 0, 1, \dots, m-1 $. Then the $ Reg.A $ will be the state \[\frac{1}{{\sqrt {{2^m}} }}\sum\limits_{k = 0}^{{2^m} - 1} {{e^{2\pi \boldsymbol {i}\theta k}}\left| k \right\rangle }. \]  Applying the inverse quantum Fourier transform, $ Reg.A $ becomes $ \ket{\tilde \theta} $, which is an approximation of $ \ket{\theta} $. After the controlled $ R_{y}(2\tilde \theta) $ rotations acting on the  ancilla $ \ket{0} $, we obtain the state
$$ Cf(\lambda)\ket{1}\ket{\lambda} + \sqrt{1-(Cf(\lambda))^{2}} \ket{0}\ket{\lambda}. $$
Finally if the ancilla is $ \ket{1} $, we have $ Cf(\lambda)\ket{\lambda} $.

\section {\label{Sec3}Some Applications}
\subsection{Applications in HHL algorithm}

\begin{figure*}[htbp]
	\centering
	\includegraphics[scale=0.7]{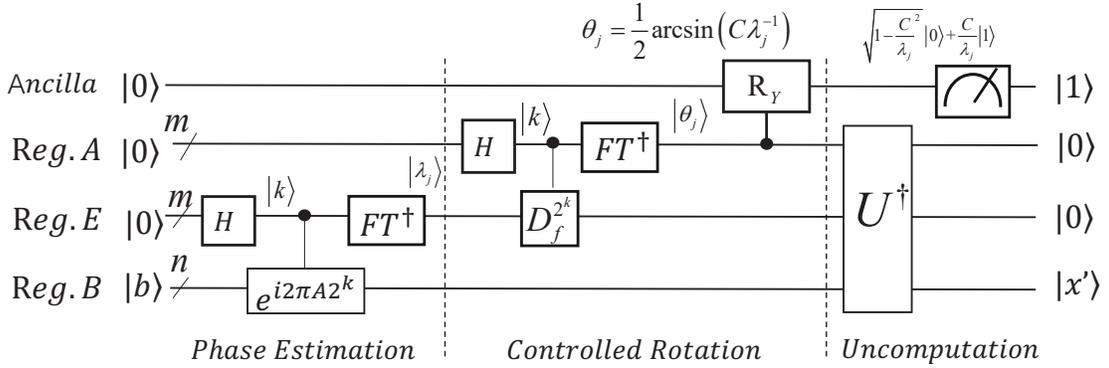}
	\caption{Overview of the quantum circuit for HHL algorithm.}
	\label{fig:hhl}
\end{figure*}

The problem of linear system can be written as 
\begin{eqnarray}
Ax = b,
\label{eq:axb}
\end{eqnarray}
where $ A \in R^{N \times N} $, $ x,b \in R^{N} , N = 2^{n}.$  $ A $ can be represented by an $ N \times N $ Hermitian matrix with a spectral decomposition of $ A = \lambda_{j}\ket{u_{j}}\bra{u_{j}} $, where $ \lambda_{j} $ is the $ j $th eigenvalue of matrix $ A $ with the corresponding eigenvector $ \ket{u_{j}} $. For convenience, we can assume that $ b $ and $ x $ are normalized and map them to the quantum states $ \ket{b} $ and $ \ket{x} $ respectively. The overview of the quantum circuit for HHL algorithm is presented in Fig.~\ref{fig:hhl} and a simple description is listed as follows. Note that we have assumed that all computations have been exact. The initial state of $ Reg.E $ and $ Reg.B $ is
\begin{eqnarray}
\ket{0}^{\otimes m}\ket{b}=\sum_{j=0}^{N-1}b_{j}\ket{0}^{\otimes m}\ket{j} = \sum_{j=0}^{N-1}\beta_{j} \ket{0}^{\otimes m} \ket{u_{j}}.
\end{eqnarray}

(i) Apply phase estimation with
\begin{eqnarray}
{e^{\boldsymbol {i}2\pi A}} = \sum\limits_{j = 0}^{N - 1} {{e^{\boldsymbol {i}2\pi{\lambda _j}}}\left| {{u_j}} \right\rangle \left\langle {{u_j}} \right|},
\end{eqnarray}
where we can assume without loss of generality that the evolution time $ t = 2\pi $. The registers become the state 
\begin{eqnarray}
\sum\limits_{j = 0}^{N - 1}{{\beta _j}\left| {{\tilde \lambda _j}} \right\rangle \left| {{u_j}} \right\rangle }. 
\end{eqnarray}
where $ \tilde \lambda _j $ is an approximation of $ \lambda _j $.

(ii) Add an ancilla qubit and perform the operation in Eq.~(\ref{U_cr}), we have 

\begin{eqnarray}
\sum\limits_{j = 0}^{N - 1} {\left( {\sqrt {1 - {{\left( {C{\tilde \lambda _j}^{ - 1}} \right)}^2}} \left| 0 \right\rangle  + C{\tilde \lambda _j}^{ - 1}\left| 1 \right\rangle } \right){\beta _j}\left| {{\tilde \lambda _j}} \right\rangle \left| {{u_j}} \right\rangle }. 
\end{eqnarray}

(iii) Apply the inverse of all operations before $ R_{y} $. If the eigenvalues are perfectly estimated, the result would be 
\begin{eqnarray}
\sum\limits_{j = 0}^{N - 1} {\left( {\sqrt {1 - {{\left( {C{\tilde \lambda _j}^{ - 1}} \right)}^2}} \left| 0 \right\rangle  + C{\tilde \lambda _j}^{ - 1}\left| 1 \right\rangle } \right){\beta _j}{{\left| {\rm{0}} \right\rangle }^{ \otimes m}}\left| {{u_j}} \right\rangle}.
\end{eqnarray}

(iv) Perform the measurement on the ancilla qubit, if the ancilla qubit is in the state $ \ket{1} $, then the register $ B $ becomes the state
\begin{eqnarray}
\left| {x^{'}} \right\rangle  = \frac{1}{{\sqrt {\sum {{{\left( {C{\tilde \lambda _j}^{ - 1}{\beta _j}} \right)}^2}} } }}\sum\limits_{j = 0}^{N - 1} {C{\tilde \lambda _j}^{ - 1}{\beta _j}\left| {{u_j}} \right\rangle },
\end{eqnarray}
where $ \left| {x^{'}} \right\rangle $ is an approximation of $ \left| x \right\rangle $. The steps (ii) can be seen as the procedure of controlled rotation implemented by the method in Sec.~\ref{Sec2}. The elements of diagonal operator $ D_{f} $ are $ e^{{2 \pi \boldsymbol {i}} \theta_{j}} $, with $  \theta_{j} = \frac{1}{2}\arcsin \left( {C{\lambda_{j} ^{ - 1}}} \right) $.

\subsection{Applications in HHL-based QML algorithms}
The HHL algorithm is of great inspiration to the design of quantum machine learning (QML) algorithms. The algorthms that utilize HHL algorithm as a subroutine or inspired by HHL algorithm can be defined as HHL-based QML algorithms. By changing the value of $ \theta $ in the diagonal element $ e^{{2 \pi \boldsymbol {i}} \theta} $ of the unitary diagonal matrix, our proposed controlled rotation can be applied to the following HHL-based QML algorithms.

\subparagraph{QSVT:}
Based on the singular value decomposition of a matrix, singular value thresholding (SVT) is a fundamental core module in computer vision and machine learning. The quantum SVT (QSVT) algorithm provides an exponential speed improvement over the classical algorithm \cite{duan2018efficient}. The QSVT algorithm involves a controlled rotation $  $ 
\begin{eqnarray}
\begin{array}{l}
U_{QSVT}: \left| 0 \right\rangle \left| \lambda_{j}  \right\rangle \\
\mapsto \left( {\frac{{\gamma \left( {\sqrt \lambda_{j}  - \tau } \right)}}{{\sqrt \lambda_{j}  }}\left| 1 \right\rangle  + \sqrt {1 - \frac{{{\gamma ^2}{{\left( {\sqrt \lambda_{j}   - \tau } \right)}^2}}}{\lambda_{j} }} \left| 0 \right\rangle } \right)\left| \lambda_{j}  \right\rangle 
\end{array}.
\label{qsvt}
\end{eqnarray}
where $ \gamma $  and $ \tau $ are constants. If the method in Sec.~\ref{Sec2} is applied to implement $ U_{QSVT} $, $ \theta_{j} $ in the corresponding diagonal operator included in the phase estimation should be $ \dfrac{1}{2}{\arcsin \left( \frac{{\gamma \left( {\sqrt \lambda_{j}   - \tau } \right)}}{{\sqrt \lambda_{j}  }} \right)} $.
\subparagraph{QAOP:} Ref.~\cite{PhysRevA.99.032311} proposed a quantum A-optimal projection (QAOP) algorithm that can be used to speed up the learning process of an important dimensionality reduction algorithm in pattern recognition and machine learning. The QAOP algorithm assumes that $ \sigma $ is the singular value of an input matrix  $ \tilde X $ and $ \beta^{(i)} $ is the singular value of the projection matrix $ A^{(i)} $, which is computed by the $ i $th iteration of the QAOP algorithm. The QAOP algorithm mainly includes phase estimation and a controlled rotation
\begin{widetext}
\begin{equation}
{U_{QAOP}}:\left| 0 \right\rangle \left| {{\sigma _j^2}} \right\rangle \left| {{{\left( {\beta _j^{\left( {i - 1} \right)}} \right)}^2}} \right\rangle  \mapsto \left[ {\rho \left( {1 + \frac{{{\lambda _2}}}{{{{\left( {{\sigma _j}\beta _j^{\left( {i - 1} \right)}} \right)}^2}}}} \right)\left| 1 \right\rangle  + \sqrt {1 - {\rho ^2}{{\left( {1 + \frac{{{\lambda _2}}}{{{{\left( {{\sigma _j}\beta _j^{\left( {i - 1} \right)}} \right)}^2}}}} \right)}^2}} \left| 0 \right\rangle } \right]\left| {\sigma _j^2} \right\rangle \left| {{{\left( {\beta _j^{\left( {i - 1} \right)}} \right)}^2}} \right\rangle,
\end{equation}
\end{widetext}
where  $ \rho < 1 $ is used for normalization of the quantum state and $\lambda_{2}$ is a small regularization coefficient. In this case, to implement $ U_{QAOP} $, $ \theta_{j} $ of the corresponding diagonal operator should be $  \dfrac{1}{2}arcsin \left( \rho \left( {1 + \frac{{{\lambda _2}}}{{{{\left( {{\sigma _j}\beta _j^{\left( {i - 1} \right)}} \right)}^2}}}} \right) \right) $.

\subparagraph{QKPCA:}
The Ref.~\cite{Li_2020} illustrated an architecture to simulate the arbitrary nonlinear kernels and further proposed the quantum kernel principal component analysis (QKPCA) algorithm, which also involved a controlled rotation 
\begin{equation}
{U_{QKPCA}}:\left| 0 \right\rangle \left| \lambda_{j}  \right\rangle  \mapsto \left( {\frac{1}{{\sqrt \lambda_{j}  }}\left| 1 \right\rangle  + \sqrt {1 - \frac{1}{\lambda_{j} }} \left| 0 \right\rangle } \right)\left| \lambda_{j}  \right\rangle,
\end{equation}
where $ \lambda_{j} $ is supposed to be larger than $ 1 $ in this situation. Here $ \theta_{j} $ of the corresponding diagnal operator prepared for $ U_{QKPCA} $ should be $ \dfrac{1}{2}arcsin\left( \frac{1}{{\sqrt \lambda_{j}}}  \right) $.

\subsection{Other Applications}

The encoding of the classical data can be maily categorized into two types: analog encoding (or called amplitude encoding), which encodes data into amplitudes of a state, and digital encoding (or called basis encoding), where the data are stored as qubit strings. The former is the basic condition for quantum acceleration, whereas the latter is required to perform arithmetics on quantum devices. Ref.~\cite{PhysRevA.99.012301} has proposed algorithms that convert these two encodings to one another, where the proposed algorithms are seriously based on the fact that the transformation 
\begin{eqnarray}
% \ket{x}\ket{0}^{\otimes m} \mapsto \ket{x}\ket{\tilde f\left( x \right)}
\ket{x}\ket{0}^{\otimes m} \mapsto \ket{x}\ket{f\left( x \right)}
\label{U_}
\end{eqnarray}
can be performed effectively. However, Ref.~\cite{PhysRevA.99.012301} still omits the details of Eq.~(\ref{U_}), which is a small but not negligible part of their proposed algorithms. The method in Sec.~\ref{Sec2} can just complement this technical detail without considering the complexity. Thus the controlled rotation module can also be applied in arbitrary nonlinear transformations of amplitudes of a quantum state, data loading and quantum state preparation \cite{PhysRevA.99.012301}. 

Moreover, Eq.~(\ref{U_}) has been used as an important part in implementing quantum neurons to merge the nonlinear, dissipative dynamics of neural computing into the linear, unitary quantum system \cite{schuld2015simulating, 8848854, PhysRevA.102.052421}. The diagonal matrix has stored a set of weights, inner products or activation function values in its diagonal elements.

\section{\label{Sec4}Experiments}

\begin{figure*}[htbp]
	\centering
	\includegraphics[scale=0.5]{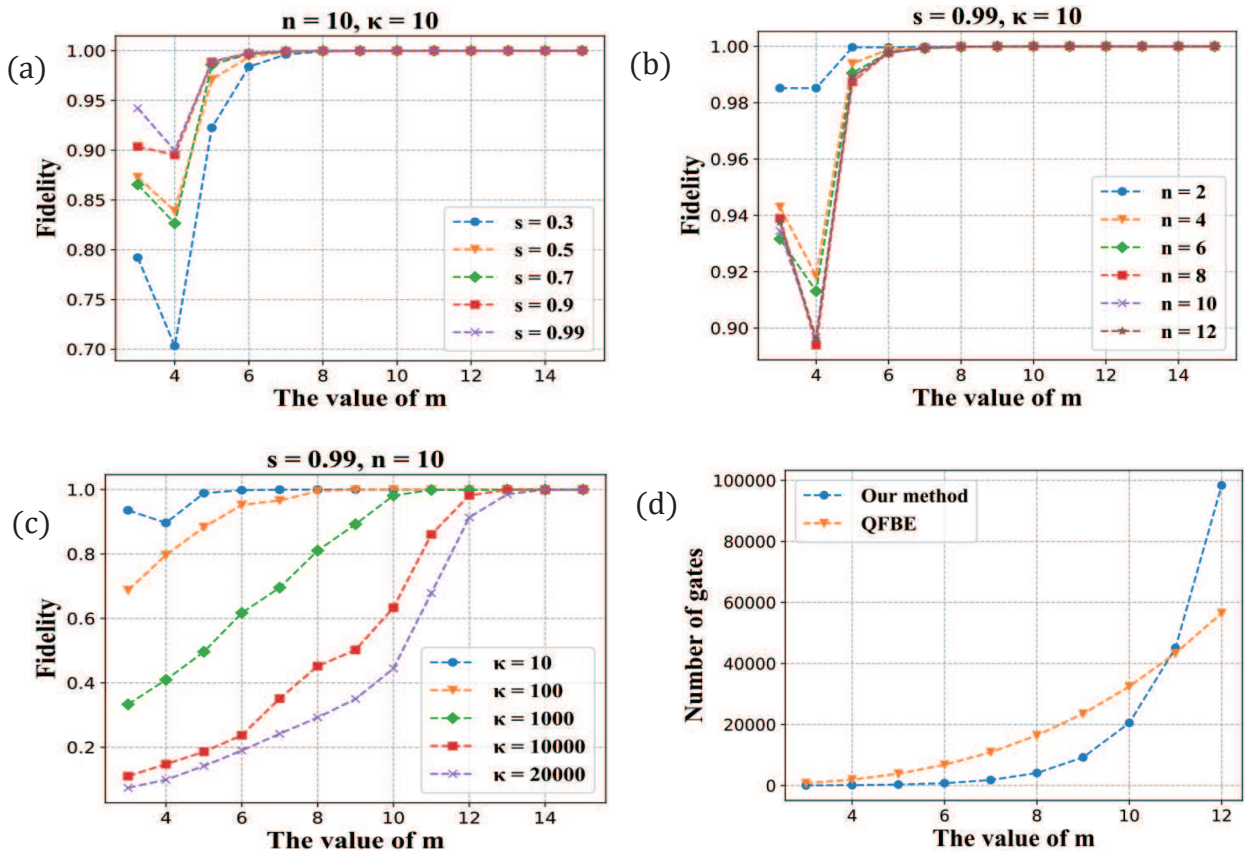}
	\caption{The results of numerical simulation.}
	\label{fig:simulation}
\end{figure*}

As the HHL algorithm is a "proof of concept", the implementation of it has become an important task. Although Refs.~\cite{PhysRevLett.110.230501, PhysRevA.89.022313, barz2014two, PhysRevLett.122.060504} has used several qubits to verify the correctness of the HHL algorithm, the stability and practicability of the HHL algorithm running on a universal quantum computer are still unknown. In this section, we will use numerical simulations to illustrate the influence of controlled rotation on the fidelity of implementing the HHL algorithm. We use the controlled variable method to explore the effect of the following four factors on the fidelity of implementing the HHL algorithm: the scaling factor $ s $, the condition number $ \kappa $ of $ A $, the dimension $ 2^{n} $ of $ A $  and the number of output qubits $ m $ . The process of simulation experiments is listed as follows.

(i) Randomly generate Hermitian matrices with uniform distribution of eigenvalues in $ (0,1] $.  Here, we have assumed all the generated Hermitian matrices are row-sparse and row-computable. To ensure the generality of simulations, the eigenvalues of each simulation are re-generated randomly.

(ii) Take $ m $ decimal places of the binary eigenvalues, then we can get an approximation $ \tilde \lambda $. In the simulation, all $ \tilde \lambda $ equal to 0 will be changed to $ \underbrace {0 \cdots 0}_{m - 1}1 $ as the object of inversion cannot be 0.

(iii) Caculate $ \arcsin \left( {C{{\tilde \lambda }^{ - 1}}} \right) $ and intercept its binary form by $ m $ bits, we have an approximation $ 2\tilde \theta \approx  2\theta = \arcsin \left( {C{{\lambda }^{ - 1}}} \right) $.

(iv) Caculate $ sin\tilde \theta $ and intercept its binary form by $ m $ bits, we obtian the simulation result $ \widetilde  {C{\lambda ^{ - 1}}} $, which is an approximation of the theoretical result $ {C{\lambda ^{ - 1}}} $.

Noting that the largest eigenvalue of $ A $ is always $ 1 $, we have $ \kappa = 1/\lambda_{min} $. Since $ C \in O(1/\kappa) $, the scaling factor $ s \in (0,1)$  satisfies $ C = s \times \lambda_{min} $. We define the fidelity of implementing the HHL algorithm as $ \bra{x^{'}} x\rangle $. 

Fig.~\ref{fig:simulation}\textcolor{blue}{(a)} shows the result of fidelity varying with $ m $ and $ s $ under the condition of $ n=10 $ and $ \kappa=10 $. We observe that as the parameter $ s $ gets closer to 1,the fidelity increases. Fig.~\ref{fig:simulation}\textcolor{blue}{(b)} shows the result of fidelity varying with $ m $ and $ n $ under the condition of $ s=0.99 $ and $ \kappa=10 $. We come to a conclusion that smaller $ n $ represents better performance in fidelity, although the improvement is subtle. Fig.~\ref{fig:simulation}\textcolor{blue}{(c)} shows the result of fidelity varying with $ m $ and $ \kappa $ under the condition of $ s=0.99 $ and $ n=10 $. We conclude that the larger the parameter $ \kappa $, the larger the number of qubits $ m $ required to maintain high fidelity.

\section {\label{Sec5}Analysis and Disscusions} 

\subsection{Complexity}

Taking the HHL algorithm as an example, we will illustrate that the controlled rotation will not influence the acceleration of the quantum algorithm. Based on phase estimation and amplitude amplification, the details of the complexity and error analysis of HHL algorithm have been shown in Ref.~\cite{PhysRevLett.103.150502}. $ e^{iAt} $ can be simulated in time $ O($ log $(N)d^{2}t) $ \cite{berry2007efficient}, where $ d $ is the sparseness of $ A $. If the phase estimation involving $ e^{iAt} $ errs by $ O(1/t) $, which translates into a relative error of $ O(1/\lambda t) $ in $ \lambda^{-1} $. The HHL algorithm has considered the complexity of the controlled rotation without a specific implementation. If $ \lambda \geq 1/ \kappa $ taking $ t = O(\kappa / \varepsilon) $ induces a final error of $ \varepsilon  $. Considering all of these above, the running time of HHL algorithm turns to $ O({{\log \left( N \right){d^2}{\kappa ^2}} \mathord{\left/{\vphantom {{\log \left( N \right){d^2}{\kappa ^2}} \varepsilon }} \right. \kern-\nulldelimiterspace} \varepsilon }) $, which offers an exponential speedup over the fastest classical algorithm: the conjugate gradient method. The conjugate gradient method runs in $ O(N\kappa) $ (or $ O(N{\sqrt {\kappa }}) $ for positive semidefinite matrices). Typically, the condition number $ \kappa $ is taken as $ O({1 \mathord{\left/{\vphantom {1 \varepsilon }} \right.\kern-\nulldelimiterspace} \varepsilon }) $, where $ \varepsilon  = {2^{ - m}} $. In this case, the running time of HHL algorithm increases exponentially with the parameter $ m $, which is one of four caveats for HHL proposed in  Ref.~\cite{aaronson2015read}. These caveats can be crucial in practice. Thus the speed-up of HHL algorithm must be based on that both $ \kappa $ and $ {1 \mathord{\left/{\vphantom {1 \varepsilon }} \right. \kern-\nulldelimiterspace} \varepsilon } $ are  poly $ \log \left( N \right) $. 

In a similar way, even if the running time of the controlled rotation is $ O(2^{m}) $, it would be translated into $ O(poly({1 \mathord{\left/{\vphantom {1 \varepsilon }} \right. \kern-\nulldelimiterspace} \varepsilon })) $. The controlled rotation would not influence the acceleration of original algorithms theoretically.

\subsection{Comparison}
In this section, we analyze the space (circuit width or the qubit complexity) and time (circuit depth or the gate complexity) resources required for different methods of the controlled rotation module. As any unitary operation can be decomposed into elementary gates \cite{PhysRevA.52.3457}, the elementary gate library is universal. We assume that any single- and two-qubit gate can be seen as elementary  gates. To analyze the gate complexity, we use elementary gate complexity. While talking about the actual value, we use elementary gates and Toffoli gates. In the previous works, the common used methods that can construct any controlled rotation are mainly divided into two types: polynomial-fitting function method and  measurement-based method. Table.~\ref{tab1} shows the comparisons of these different methods.

The polynomial-fitting function method includes Newton iteration,Taylor series, quantum function value binary expansion (QFBE) method, and so on. The basic principle of the polynomial-fitting function method is to transform the function calculation into quantum version of multiplication and addition, which has been studied in     Refs.~\cite{saeedi2013synthesis,PhysRevA.54.147,PhysRevA.54.1034,841192,draper2000addition,cuccaro2004new,10.5555/2011670.2011672,10.5555/2012086.2012090,_lvarez_S_nchez_2008,10.5555/2011464.2011476,rieffel2011quantum}. Reversibility of any unitary operation effects that addition and multiplication cannot be directly deduced from their classical Boolean counterparts. Ref.~\cite{PhysRevA.54.147} first showed that the original addition of two registers $ \ket{a} $ and $ \ket{b} $ can be written as $ \left| {a,b,0} \right\rangle  \mapsto \left| {a,b,a + b} \right\rangle $ or $ \left| {a,b} \right\rangle  \mapsto \left| {a,a + b} \right\rangle  $. If both $ a $ and $ b $ are encoded on $ m $ qubits, implementing an adder costs $ 3m+1 $ qubits, $ 20m+2 $ CNOT gates and $ 20m-10 $ Toffoli gates. Based on adder, multiplier is computationally more expensive, where a multiplier of $ a $ with $ b $ can be thought of as adding $ a $ to itself $ b $ times. In this way, multiplier takes more qubits and quantum gates to complete the multiplicative task. Thus the polynomial-fitting function method is not a resource-saving method at some times. Recently, Ref.~\cite{wang2020quantum} develops the QFBE method to evaluate the transcendental functions in a recursive way. Totally, the QFBE method uses $ 34{m^3} - 16{m^2} + 4m $ quantum gates to implement the controlled rotaion.

The measurement-based method can be divided into two types: one is the RUS arithmetic (or function synthesis) \cite{10.5555/3179320.3179329} and the other is classical computing \cite{lee2019hybrid}. Focused on optimizing space, RUS arithmetic encodes numbers in the amplitudes of a qubit, or more properly as polar angles on the Bloch sphere. The core idea is utilizing measurement to implement nonlinear mappings between sets of input and output rotation angles. However, RUS arithmetic has not discussed an important problem that how many repetitions of these circuits are needed before a successful result is observed with high probability. Classical computing directly analyzes measurement outcomes from $ \ket{\lambda} $ by means of classical computers. Based on the analyzed data, a simpler circuit implementation of the original part can be built together with reducing the circuit depth and space. Similarly, classical computing does not solve the problem of the number of measurements. The measurement-based method generally requires polynomial time and space complexity.

The idea of our method in Sec.~\ref{Sec2} is fully combining the function outputs and quantum circuit design. Containing $ 2^{m} $ parameters, the accurate evaluation of diagonal unitary operations might be the most resourceintensive element. For space optimizing, we need to use the decomposition method to realize the diagonal operation $ D $.  Alternating controlled not gates and $ R_{z} $ rotations, the best-known compiling algorithm provided circuits of size $ 2^{m} -3 $ for arbitrary $ m $-qubit diagonal unitaries \cite{DBLP:journals/qic/BullockM04}. As the result of the power exponent of the diagonal matrix is also diagonal, $ 2^{m} -3 $ Toffoli gates and controlled-$ R_{z} $ gates are also enough for implementing controlled $ D^{2^{i}} $ gates. An efficient implementation of the phase estimation includes $ O(m^{2}) $ elementary gates for an inverse quantum Fourier transform and one call to controlled $ D^{2^{i}} $. That is it totally needs $ m\left( {{2^{m + 1}} - 3} \right) + m\left( {m + 1} \right)/2 + m $ single- and two-qubit gates and Toffoli gates.

\begin{table}[htbp]
	\caption{Comparisons of different methods.} 
	\centering
	\addtolength{\tabcolsep}{0.8mm} % 控制列间距
	\setlength{\tabcolsep}{0.8mm}{
		\begin{tabular}{ccc}
			% \toprule[0.75pt]
			\hline \hline
			&  \multicolumn{1}{c}{\begin{tabular}[c]{@{}c@{}} Qubit complexity \end{tabular}}& \multicolumn{1}{c}{\begin{tabular}[c]{@{}c@{}} Gate complexity \end{tabular}}   \\ \hline
			\midrule
			Newton iteration  & $ O(m^{3}) $ & $ O(m^{4}) $  \\ 
			QFBE  & $ O(m^{2})  $ & $ O(m^{3}) $  \\  
			Measurement-based   & $ O(poly(m)) $ & $ O(poly(m)) $  \\ 
		%	Look-Up Table  & $ O(m) $ & $ O(m^{3}2^{m}) $  \\ 
			Our method & $ O(m) $ & $ O(m2^{m}) $  \\  
			\hline \hline
			% \bottomrule[0.75pt]
	\end{tabular}}
	\label{tab1}
\end{table}

Fig.~\ref{fig:simulation}\textcolor{blue}{(d)} shows the total number of quantum gates that are needed to implement the controlled rotation in HHL algorithm with two different methods.  When $ m $ is less than 11, the number of quantum gates required by our method is less than QFBE. At this time, the HHL algorithm using our method can already make the fidelity of inversing the matrices with $ \kappa \leq 1000 $ close to 1, where the condition number 1000 already contains most of the matrices that need to be inverted in practice. Thus, our method only requires the least ancillas or the least quantum gates to maintain the high fidelity of quantum algorithms.

We define the diagonal unitary operator as
\begin{eqnarray}
D = diag\left\{ {{e^{2\pi \boldsymbol {i} \cdot {\theta _0}}}, {{e^{2\pi \boldsymbol {i} \cdot {\theta _1}}}}, \cdots , {e^{2\pi \boldsymbol {i} \cdot {\theta _{{2^m} - 1}}}}} \right\}, 
\end{eqnarray}
where $ \theta_i \in (0,1]$ and $ i \in \{0,1,\cdots,2^{m}-1 \} $. Supposing the values of all $ \theta_i $ are known in order, then the controlled-$ D $ is written as
\begin{eqnarray}
CD = C^{m}R_{z}(\theta _0) \otimes \cdots \otimes C^{m}R_{z}(\theta _{2^{m}-1}).
\end{eqnarray}
That is the $ CD  $ can be implemented with $2^{m}$ different $C^{m}R_{z} $ gates, where a $ C^{m}R_{z} $ gate can be simulated by $ O(m) $ gates with one auxiliary qubit \cite{PhysRevA.52.3457}. If the number of different $ \theta _i $ is $ O(poly(m)) $, the gate complexity would be $ O(poly(m)) $. In this way, our method would be more competitive.

\subsection{The Threat of Oracle Expansion}
Quantum oracle, or called black box, is an important tool used in quantum computation. Many quantum algorithms would like to use the oracle as a part of themselves, such as phase flip of target state in Grover's algorithm, modular exponentiation in Shor's algorithm and contrlled rotation in HHL algorithm. The focuses of designing a quantum algorithm are complexity and error analysis. Most papers deal with a quantum algorithm but omit the details of implementation. Each oracle is implemented by a quantum algorithm itself. When an oracle is used to implemente an algorithm, it often significantly contributes to the depth and width of the corresponding algorithm: the depth of an algorithm is the number of gates to be performed sequentially, and its width is the number of qubits it actually manipulates. Although oracles are usually considered effective, they may affect the efficiency of implementing the algorithm within limited conditions. To evaluate whether a gate-based algorithm can be implemented and executed on a particular NISQ device, we have to carefully consider the exact value and complexity of the width and depth of a circuit.

Ref.~\cite{leymann2020bitter} has discussed different impact factors of algorithm implementations and drawn a conclusion: a quantum algorithm using oracles (i.e. black box functions in general) must be modified by expanding these functions as quantum circuits themselves and substituting these functions by the corresponding circuits. For modular gate-based algorithms, each module will affect the implementation of the entire algorithm. Existing generic constructions focus on the complexity of the circuit depth rather than actual value. In theory, the polynomial algorithm is better than the exponential algorithm. However, time cost of the exponential algorithm in a certain range will be less, when the factor of the polynomial is large. The result in Fig.~\ref{fig:simulation}\textcolor{blue}{(d)} validates our point. 

Although our method allows to reduce the quantum resources in some way, the resulting circuits are still quite expensive, especially in terms of the number of required quantum gates. At present, there is no satisfactory method to realize arbitrary controlled rotation, then the acceleration of some quantum algorithm may be inhibited.

\section{\label{Sec6}Conclusions}
In conclusion, our work implements arbitrary controlled rotation module for HHL algorithm. The proposed method can also be applied in HHL-based quantum algorithms and quantum machine learning. The analysis show that the method theoretically makes the quantum algorithms have high fidelity without influencing the acceleration of original algorithms. Compared with the polynomial-fitting function method or some other methods, our method only requires the least ancillas or the least quantum gates in some way. Numerical simulations illustrate the effectiveness of the proposed method. If the corresponding diagonal unitary matrix can be effectively decomposed, the method is also polynomial in time cost. We highlight a fact that there is still no satisfactory method to realize arbitrary controlled rotation. We also hope that there can be more researches in the implementation of quantum algorithms to further reduce the cost of arbitrary controlled rotation. In the future, we will further optimize the quantum circuit to make the gate-based quantum algorithms more forcefully demonstrate its superiority.

\begin{acknowledgments}
 This work was supported by the National Natural Science Foundation of China under Grant 61873317  and in part by the  Guangdong Basic and Applied Basic Research Foundation under Grant 2020A1515011375.
\end{acknowledgments}

\label{sec:TeXbooks} 

%\bibliography{Controlled-rotation}% Produces the bibliography via BibTeX.
%\begin{thebibliography}{10}
%\end{thebibliography}

\end{document}